# Phonons in Crystals using Inelastic X-Ray Scattering

Alfred Q.R. Baron[*]

*Materials Dynamics Laboratory, RIKEN SPring-8 Center, 1-1-1 Kouto, Sayo, Hyogo, 679-5148*
*Research and Utilization Division, SPring-8/JASRI, 1-1-1 Kouto, Sayo, Hyogo, 679-5198*

The use of inelastic x-ray scattering (IXS) to investigate phonons and phonon dispersion in crystals opens the field of phonon spectroscopy, allowing measurements on tiny samples: micrograms of material, or single crystals ~10 to 100 μm diameter are sufficient. Here we review the technique as it is practiced at BL35XU of SPring-8, in Hyogo prefecture. An introduction is provided that is aimed at potential IXS users, highlighting the unique and advantageous aspects of the technique. Present work, including several example experiments, and future directions are discussed.

## 1. Introduction & Scope

The study of lattice excitations, phonons, in crystalline materials has a long and rich history. While a variety of indirect methods were initially used to verify the phonon model, the direct observation of phonon dispersion using inelastic neutron scattering by Brockhouse and others [1] opened the field of phonon spectroscopy as a momentum and energy resolved technique, allowing full exploration of phonon dispersion. It was tacitly recognized that x-rays might also be used to probe dynamics via thermal diffuse scattering, but largely ignored due to the fundamental mismatch in energy scales: Å wavelength x-rays have energies ~10 keV, so resolving ~meV phonon excitations is not obvious.[+]

The idea of using inelastic x-ray scattering as a tool for measuring phonon dispersion was discussed as early as 1980 [2] and this was followed by a published proposal [3], and then construction and demonstration of the first inelastic x-ray scattering (IXS) spectrometer at HASYLAB in Hamburg, Germany[4]. Since then, especially with the development of 3rd generation sources, and the success of the first IXS beamline at the European Synchrotron Radiation Facility (ESRF) [5], there has been sustained effort in the field, and meV-resolved IXS has come into its own as a technique.

The present paper provides an introduction to phonon measurements by inelastic x-ray scattering (IXS). It is based primarily on the author's experience at BL35XU of SPring-8 and in the RIKEN Materials Dynamics Laboratory [6][7]. It is aimed primarily at people who might be interested in entering the field. We place the IXS technique within the context of both alternative methods of investigation and current physical questions. Instrumentation is only mentioned in so far as it impacts the types of measurements possible. To be precise: the focus of this paper is non-resonant, ~meV resolved, inelastic x-ray scattering as probe of atomic dynamics in crystals.

There are now 5 facilities world-wide doing these types of measurements, including two beamlines at the European Synchrotron Radiation Facility (ESRF) in Grenoble France [8] two at the Advanced Photon Source (APS) in Argonne, Illinois, in the US [9], and BL35XU of SPring-8 in Harima Science

---

[*] baron@spring8.or.jp
[+] Reference [12] contains a more detailed historical summary of phonon investigations.

Garden City of Hyogo Prefecture, Japan [10]. Additional beamlines have been funded including a uniquely powerful one using a long undulator (BL43LXU) through RIKEN in Japan, expected to be operational in 2012, as well as a beamline using a new and, as yet, unproven, method at the NSLS-II, which will see first light in 2014. All presently operating facilities are highly oversubscribed.

There is extensive literature available discussing phonons in crystals: the early book by Born and Huang [11] remains pertinent and useful, while more recent reviews include the several volume set by Breusch [12], or the shorter text by Srivastava [13]. A comprehensive collection of papers can be found in the books edited by Horton and Maradudin [14]. The present author has also found the book by Dorner [15] to be a useful practical discussion about measuring phonons in crystals. Introductions to scattering theory, as related to phonons, can be found in the books by Squires [16] and by Lovesey [17], as well as the paper by van Hove [18], with explicit formulation for x-rays in [19]. Specific reviews focusing on high resolution inelastic x-ray scattering include [20, 21, 22]. One also notes the reviews, in Japanese, focusing on instrumentation at BL35XU [23], and on liquid scattering [24]. A discussion of much of the instrumentation and results specific to BL35XU of SPring-8 can be found in the 5 year report [25].

## 2. X-Rays Compared to Neutrons for Phonon Measurements

Inelastic neutron scattering (INS) has long been the method of choice for measuring phonon dispersion. It is therefore worthwhile to compare IXS and INS. The most fundamental difference is that IXS probes the coherent motion of the electronic cloud around atoms (Thomson scattering) while INS probes the nuclear motion. The equivalence of the two relates to the "adiabatic approximation", that the electrons move with the nuclei. This is generally an excellent approximation, especially as the strongest x-ray scattering at the momentum transfers of interest comes from the core electrons. The author is not aware of any cases of meV-resolved IXS experiments looking at phonons where failure of the adiabatic approximation has led to significant effects.

There are a variety of other differences between the two methods that originate from both practical and fundamental considerations. Specific features of IXS include

1. Decoupling of energy and momentum transfer.
2. Access to small, ~microgram, samples.
3. Energy resolution that is independent of the energy transfer.
4. Nearly no intrinsic backgrounds.
5. Simple and good momentum resolution.

The first two advantages are probably the most significant, and, perhaps unsurprisingly, can be considered as resulting from precisely the technical challenges associated with IXS. The first (also the third and fifth) follows from the fact that x-ray energies (~20 keV) are much larger than the measured energy transfers (~1 to 200 meV), unlike neutrons where the probe energy is often very similar to the phonon energy, ~20 meV. It is especially important in liquid experiments where one would like access to large energy transfer at small momentum transfers, and is less important for crystalline materials where one can often work in higher Brillouin zones. The access to small samples, (2) above, is perhaps the most important in the present discussion. It follows from the very high flux and brilliance of synchrotron radiation sources, with the option to focus beams easily to ~100 microns in diameter, and, with some losses, to ~10 microns. It means that one can investigate small samples, including samples at very high (e.g. earth's core) pressures in diamond anvil cells, and small crystals of new materials. The latter is especially important in the world of modern materials synthesis, where often the most interesting materials are only available in small very single crystals – for example it has now been about 7 years since $MgB_2$ was shown to have a very high $T_c$, and available crystals are still too small for measurements of phonon dispersion by neutron scattering. Thus, the huge Kohn anomaly associated with the electron phonon coupling that drives the high-$T_c$ of this material has only been directly probed by x-ray scattering [26].

The other items also play important roles, depending on the experiment. The independence of energy resolution and transfer means that one can investigate high-energy (say 100 meV) phonon modes with high (~ 1.5 meV) resolution, whereas neutrons typically have 5 to 10% energy resolution. The low background in IXS work stems from the fact that x-ray attenuation lengths in most materials are so short as to prevent multiple scattering, and that incoherent scattering with x-rays is at much higher energy scales (>eV) due to the small electron mass. Thus, the only background appearing at phonon energy scales for x-ray scattering is the multi-phonon contribution. One should note, however, that the resolution function in IXS is fundamentally Lorentzian, so one needs to be careful about tails from elastic scattering, which can extend to large energy transfer and create an effective background. The momentum transfer resolution is determined by slits in front of the analyzer, and has a simple shape, independent of energy transfer. It can be reduced nearly arbitrarily (count-rates permitting) down to the level of ~0.2 nm$^{-1}$, though below that one needs to care about the detailed instrument configuration.

Neutrons remain advantageous when high-energy resolution is needed, as backscattering spectrometers provide sub-meV resolution, at least for smaller energy transfers. In addition, whereas the energy resolution is approximately Lorentzian with x-rays (as it is usually limited by dynamical diffraction) and so has long tails, that with neutrons often has shorter tails. This can make neutrons advantageous for observing weak modes near to stronger ones, or for measuring modes in the presence of strong elastic backgrounds. Neutrons remain extremely competitive when large single crystals of heavier materials are available, whereas, x-rays are limited by the short penetration length into the sample. Also, advanced neutrons spectrometers allow collection of a huge swath of momentum space at one time. As the relationship of the isotopic species and cross-section for neutrons scattering is complex, there also can be advantages for each (x-rays or neutrons) depending on the atoms involved in the phonon modes of interest.

## 3. IXS Compared to Other Methods using X-Rays

There are several other techniques that may be used for investigating atomic dynamics with x-rays. On a meV scale, several of these rely on scattering from low-lying narrow (typically neV to ueV bandwidth) nuclear resonances [27, 28] in materials such as Fe, Eu, Sn, Sm, Dy, Kr, and Tm to name a few of the more accessible resonances (for a review of the larger nuclear resonant scattering field see [29]). The most robust of these techniques, sometimes called Nuclear Inelastic Scattering (NIS), uses a ~meV resolved beam incident on a sample containing the resonant material [27]. Scanning the energy of this beam about the nuclear transition energy, and selecting only the nuclear scattered events, yields the distribution of excitations in the sample that can make up the difference between the incident beam energy and the nuclear resonance energy. This is fundamentally an absorption or incoherent scattering measurement (with momentum transfer equal to the incident photon momentum), so it yields the partial density of states of resonant nuclear motion. While the method is limited to probing the motions of the resonant atom, the data is generally of much higher quality than the DOS provided by neutron scattering. The nuclear and element-selectivity can also be useful to focus on motions of particular atomic species, and the large nuclear cross-section can allow access to relatively small samples (e.g. thin layers). However, NIS does not allow measurement of phonon dispersion. Also, selection of specific phonon modes is not possible using NIS, unless they are isolated from all other modes in energy. An alternative to the NIS method, which is not limited to resonant atoms in the sample, is to use a nuclear analyzer (NA) [28]. This is similar in concept to the IXS technique discussed in the bulk of this paper, except that nuclear scattering is used to analyze the scattered radiation. This method allows simultaneous measurement of a large fraction of momentum space, or large solid angle, as determined primarily by the detector. However, it is limited because the effective bandwidth of the nuclear analyzer, is ~1 microvolt, or less. Thus it is probably most interesting when large solid angles can be collected, or when extremely high resolution is absolutely required.

Thermal diffuse scattering (TDS) is another potential probe of atomic dynamics: momentum resolved but energy-integrated measurements generally shows structure due, in part, to the phonon polarization vectors, and in part due to the 1/E scaling of phonon scattering intensities. Some information can be derived from these measurements [30] by fitting results from simple samples with models, and/or through temperature dependent measurements. However, due to the energy-integrated nature of TDS measurements, it is of limited value as a direct probe of phonons in more complex materials. TDS is potentially most useful as a complementary technique to IXS - it might be used to investigate large parts of momentum space to, for example, pinpoint changes in intensity across a phase transition. Such a TDS investigation could then be followed by a detailed investigation using IXS.

## 4. The Phonon Cross Section

The cross-section for non-resonant scattering from a material is usually divided into the product of a term relating to the probe-sample interaction and the dynamic structure factor, $S(\mathbf{Q},\omega)$, which is more directly related to atomic motions in the sample. For scattering of a photon from an initial state given by photon momentum and (complex) polarization $\mathbf{k}_1$, $\varepsilon_1$ to a final state given by $\mathbf{k}_2$, $\varepsilon_2$ the cross-section is (see references [16-19] but do note [17] only considers monatomic samples)

$$\left(\frac{d^2\sigma}{d\Omega dE}\right)_{\mathbf{k}_1\varepsilon_1 \to \mathbf{k}_2\varepsilon_2} = \frac{k_2}{k_1} r_e^2 \left|\varepsilon_1^* \bullet \varepsilon_2\right|^2 S(\mathbf{Q},\omega) \quad (1)$$

where assuming non-resonant scattering process allows us to concentrate on the energy transfer $E \equiv \hbar\omega \equiv \hbar(\omega_1 - \omega_2)$ and assuming translation invariance of the sample, allow us to focus on the momentum transfer. For photons, one has the usual relations between energy, wavelength, and momentum, $E_i \equiv \hbar\omega_i \equiv \hbar c|\mathbf{k}_i| = hc/\lambda_i$. The over-all scale is determined by the Thomson scattering factor $r_e^2 \left|\varepsilon_1^* \bullet \varepsilon_2\right|^2$, where $r_e \equiv e^2/m_e c^2 \approx 2.818$ *fm* is the usual classical radius of the electron. Note that for the non-resonant scattering discussed in this paper, the *photon* polarization enters as an un-interesting scale factor.

The dynamic structure factor is most generally written as the Fourier transform of the density-correlation operator, but, for crystalline materials, this is usually simplified by extracting the Debye-Waller factor, and expanding in a power series assuming small atomic displacements (see [16-19]). The terms of the series correspond to elastic scattering (Bragg reflections), one-phonon contributions, two phonon contributions, higher order contributions, and possible interference effects [31]. The one-phonon term is generally largest except at Bragg reflections (note that in a hypothetical extended perfect crystal, the elastic scattering is *only* at Bragg reflections). Assuming the single scattering limit (Born approximation) the coherent one-phonon scattering for crystalline materials may be written

$$S(\mathbf{Q},\omega)_{1p} = N \sum_{\substack{\mathbf{q} \\ 1st\ Zone}} \sum_{\substack{j \\ 3rModes}} \left| \sum_{\substack{d \\ Atoms/Cell}} \frac{f_d(\mathbf{Q})}{\sqrt{2M_d}} e^{-W_d(\mathbf{Q})} \mathbf{Q}\bullet\mathbf{e}_{\mathbf{q}jd} e^{i\mathbf{Q}\bullet\mathbf{x}_d} \right|^2 \delta_{(\mathbf{Q}-\mathbf{q})\tau} F_{\mathbf{q}j}(\omega) \quad (2)$$

where N is the number of unit cells illuminated by the x-ray beam, $\mathbf{q}$ is the reduced momentum transfer, within the first Brillouin zone, as determined by choosing $\tau$ to be the nearest Bragg vector to $\mathbf{Q}$, $\mathbf{Q} = \mathbf{q} + \tau$, d is an index that runs over the r atoms in the primitive cell located at $\mathbf{x}_d$, and j is the phonon mode index that runs over the 3r modes expected at any momentum $\mathbf{q}$. The x-ray form factor $f_d(\mathbf{Q})$ can be taken as nearly the free-atom result, especially as, at the momentum transfers usually of interest in phonon measurements, most of the scattering is by core electrons. One should note that $f_d(\mathbf{Q})$ does drop of rapidly with increasing momentum transfer. Values for $f_d(\mathbf{Q})$ may be found in, e.g., ref [32], and electronic tabulations are available from several sources. For any given q, there are 3r modes that can be excited. However, the intensity of each mode depends (sensitively) on the full momentum transfer Q and details of the mode motion as described by the complex phonon polarization $\mathbf{e}_{\mathbf{q}jd}$, which indicates the direction of and phasing of the motion of each of the atoms. The polarization eigenvectors are normalized, $\sum_d \mathbf{e}_{\mathbf{q}jd}^* \bullet \mathbf{e}_{\mathbf{q}j'd} = \delta_{jj'}$ and satisfy $\sum_{\substack{j \\ 3rModes}} \left(\mathbf{e}_{\mathbf{q}jd}^* \bullet \hat{\mathbf{x}}_\alpha\right)\left(\mathbf{e}_{\mathbf{q}jd'} \bullet \hat{\mathbf{x}}_\beta\right) = \delta_{\alpha\beta} \delta_{dd'}$ where $\hat{\mathbf{x}}_\alpha$ ($\alpha$=1,2,3) are orthonormal vectors. It is worth noting that the phasing of the eigenvector is subject to different conventions, and sometimes the factor, $e^{i\mathbf{Q}\bullet\mathbf{x}_d}$, is subsumed into the polarization. Further discussion can be found in references 11-15.

Phonon intensities, while often well described by eqn. (2), are not simple. Perhaps the only generally reliable statement is that long-wavelength acoustic modes are strongest near strong Bragg reflections – this follows because, for acoustic modes at small $\mathbf{q}$, $\mathbf{e}_{\mathbf{q}jd}/\sqrt{M_d}$ is approximately independent of the atom, d, (see chapter 5 of ref. 11 and chapter 1 of ref 14) so $\mathbf{Q}\bullet\mathbf{e}_{\mathbf{q}jd}/\sqrt{M_d}$ can be factored out of the sum, making S nearly proportional to the structure factor of the Bragg reflection. In simple materials, a similar argument suggests that optic modes (where different atoms in the unit cell move out of phase) are strongest in the vicinity of weaker Bragg peaks. However, as soon as the material becomes more complex, or one moves farther from the zone center, the intensity distribution can become very complicated. Figure 1 illustrates this with a calculation of a c-axis polarized arsenic breathing mode in a 1111 type Fe pnictide superconductor [33]. Considering this figure, and eqn. 2 in general, shows careful calculations and a clear picture of mode eigenvectors is necessary – otherwise it is all too easy to measure in a zone where a mode of interest is either weak, or masked by other modes.

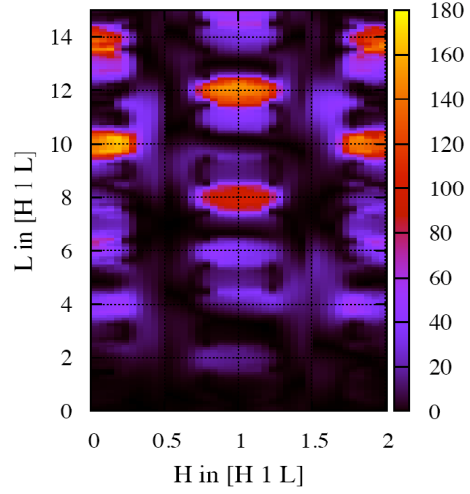

**Fig. 1.** Intensity map of a c-axis polarized As breathing mode based on a calculation of an Fe-pnictide material [33].

A comment on nomenclature is appropriate here. In particular, the terminology longitudinal and transverse is often used to describe phonon *modes* and/or measurement *geometries*. For *modes*, this refers to the directions of atomic motion relative to the propagation direction, **q**. For example, an acoustic shear mode is transverse, while the compression mode is longitudinal. For *measurement geometries* the terminology refers to direction of the reduced momentum transfer, **q**, relative to the total momentum transfer, with a longitudinal geometry having **q** parallel to **τ**, or a transverse geometry having **q** nearly perpendicular to **τ**. Certainly, if one wants to observe a transverse phonon mode, a transverse geometry is preferable, and, likewise a longitudinal geometry for a longitudinal mode. However, some care is needed because not all phonon modes are easily classified as longitudinal or transverse. Especially, as one moves away from the center of the Brillouin zone, or as materials become complex, or low dimensional, or if one moves off of high-symmetry directions, the concept of longitudinal and transverse modes can be ill-defined. In some cases (as above, fig. 1) it can be clearer to refer to modes by their eigenvectors (e.g. plane polarized, c-axis polarized), without reference to the direction of propagation.

Returning to the details of eqn (2), the Debye-Waller Factor is $e^{-2W_d(\mathbf{Q})}$ where

$$W_d(\mathbf{Q}) \;=\; \frac{\hbar}{4M_d}\frac{1}{N_q}\sum_{\substack{\mathbf{q}\\ \text{1st Zone}}}\sum_{\substack{j\\ \text{3r Modes}}}\frac{|\mathbf{Q}\cdot\mathbf{e}_{\mathbf{q}jd}|^2}{\omega_{\mathbf{q}j}}\coth(\hbar\omega_{\mathbf{q}j}\beta/2) \qquad (3)$$

$N_q$ is the number of q points in the sum and $\beta = 1/k_B T$ with $k_B \approx 0.0862$ meV/K the Boltzman constant. In many cases, $W_d(\mathbf{Q})$ is nearly atom-independent, so that it can be factored out of the sum and (mostly) ignored, the more so because calculating W requires integration over momentum space and can be slow. However, there are cases when neglecting the atom-dependence of W can be dangerous, as can occur when there are soft optical modes in the system, or when there is strong anisotropy in elastic properties, or if one approaches the Debye temperature. For example, in the case of CaAlSi, discussed below, the presence of a soft mode introduces large differences in the Debye-Waller factor for different atoms, with factors of two easily possible at room temperature [34].

The final term in eqn (2) describes the spectral shape and the temperature dependent intensity of the phonon mode. For the simplest case of harmonic phonons it is given by [16-19]

$$F_{\mathbf{q}j}^{Harmonic}(\omega) \;=\; \frac{1}{\omega_{\mathbf{q}j}}\Big[\langle n_{\omega_{\mathbf{q}j}}+1\rangle\,\delta(\omega-\omega_{\mathbf{q}j}) \;+\; \langle n_{\omega_{\mathbf{q}j}}\rangle\,\delta(\omega+\omega_{\mathbf{q}j})\Big] \qquad (4)$$

with $\langle n_{\mathbf{q}j}\rangle$ being the usual Bose occupation factor for thermal equilibrium, $\langle n_{\mathbf{q}j}+1\rangle = \left(1 - e^{-\hbar\omega_{\mathbf{q}j}/k_B T}\right)^{-1}$. Note that positive energy transfer, $\hbar\omega > 0$, is taken to indicate transfer of energy from the x-ray beam to the sample, or Stokes scattering.

For the simplest next-order approximation, one can take the spectral function to have finite line-width due to electron-phonon coupling or anharmonic phonon-phonon scattering. A reasonable approximation in many cases has the form of a damped harmonic oscillator (DHO) [35]

$$F_{\mathbf{q}j}^{DHO}(\omega) = \frac{\omega}{1-e^{-\hbar\omega/k_BT}} \frac{4}{\pi} \frac{\gamma_{\mathbf{q}j}}{\left(\omega^2 - \Omega_{\mathbf{q}j}^2\right)^2 + 4\omega^2\gamma_{\mathbf{q}j}^2} \tag{5}$$

where $\gamma_{\mathbf{q}j}$ is the line-width (half-width at half-maximum for $\gamma_{\mathbf{q}j} \ll \omega_{\mathbf{q}j}$) of the phonon mode and $\Omega_{\mathbf{q}j}^2 = \omega_{\mathbf{q}j}^2 + \gamma_{\mathbf{q}j}^2$ is the effective frequency of the mode. Equation (5) reduces to eqn (4) in the limit of small $\gamma_{\mathbf{q}j}$. This form is also used to describe excitations in disordered materials. More generally, the frequency shift and line-width, corresponding to the real and imaginary parts of the self-energy, can be frequency dependent [12][36], and higher order terms can, in principle, result in interference affects altering the line-shape [31].

The terms of higher order than the one-phonon result generally involve integrals of one-phonon parameters over the Brillouin zone. There is, therefore, a tendency to regard these contributions as not being very structured. In neutron scattering this is often reasonable, as there are also contributions from incoherent and multiple scattering, and the total of all of these can be mostly featureless, and only weakly momentum and energy dependent. However, for x-rays, one has *only* the multi-phonon background and this, in itself, is usually not flat, and can even lead to peaks in the data, as seen in [37]. While typically weaker than strong one-phonon peaks, two-phonon scattering can still make interpretation of the spectra difficult. One notes that multi-phonon contributions typically scale as a high power of the momentum transfer (i.e.: two-phonon contributions as $Q^4$) and will generally be larger at high temperature, though they can *not* be completely removed by going to low temperature.

## 5. Some Experimental Considerations

Most presently operating spectrometers for IXS are built along similar lines, employing high-order Bragg reflections in perfect silicon crystals to attain sufficient energy resolution. Typical parameters for the instrument at BL35XU of SPring-8 are given in table 1 below – resolutions between 6 and 1 meV are possible, with flux decreasing as resolution is improved. Note that at 1 meV resolution, geometric contributions (as opposed to the intrinsic rocking-curve width) have become dominant. The essential arrangement is analogous to a neutron triple-axis instrument with one "axis" for the monochromator, defining the incident beam, one axis (really a complex goniometer setup) for defining the sample orientation, and one axis for the "analyzer" defining the momentum resolution and energy resolution for the scattered beam. The location of the analyzer relative to the sample determines the momentum transfer. For those with a neutron scattering background, IXS can be considered similar to *elastic* neutron scattering (but with an energy variable) in the sense that there is no coupling of energy and momentum transfer. IXS spectrometers are generally very large (~10m analyzer arms) to reduce geometric contributions to the resolution, but, even so, it is difficult to achieve better than ~1 meV energy resolution. This is the result of the increased geometric contributions, steadily dropping flux and also because of the reductions in efficiency of silicon optics for very high order reflections at high energy – the penetration into the silicon on the Bragg reflection (the extinction length) becomes comparable to the absorption length and the reflections become weak.

| Energy (keV) | Silicon Reflection | Darwin Width (meV) | Energy Resolution (meV) |
|---|---|---|---|
| 15.816 | (8 8 8) | 4.1 | 6.0 |
| 17.794 | (9 9 9) | 1.8 | 3.0 |
| 21.747 | (11 11 11) | 0.8 | 1.5 |
| 25.702 | (13 13 13) | 0.3 | 1.0 |

**Table 1. Operating parameters of BL35XU.** Note the Darwin Width is the expected resolution for a single crystal illuminated by a plane wave, and is calculated only. The resolution is what has been measured and is given for typical operating conditions - it can be improved slightly in selected analyzer crystals. The relative flux on the sample depends on the details of the setup but is typically about 8:2:1:0.2, where one unit is about 4 GHz.

Most IXS experiments remain flux limited. In the context of synchrotron radiation experiments, it is worth emphasizing that the limiting factor in IXS spectrometer performance is usually x-ray flux (most conveniently given in units photons/s/meV) not source brilliance (photons/s/meV/source-size/source-divergence). While brilliance is required for these spectrometers, generally once a threshold level is reached to satisfy constraints of the optics, flux becomes the dominant parameter limiting spectrometer capability. Thus, all instruments presently in use employ an array of analyzer crystals, collecting data simultaneously at several momentum transfers. First introduced at the ESRF, a linear array of analyzers in the scattering plane allows immediate parallelization of data collection for disordered materials where the only free parameter is the magnitude of the momentum transfer. For crystals such an array also can allow collection of data in high-symmetry directions for a few momentum transfers for longitudinal modes. A two-dimensional array, as now exists only at BL35XU, allows a similar parallelization for transverse modes [38].

Given the flux limitations, one should design sample holders and environments to both maximize the transmitted radiation and minimize backgrounds. Air scatter can be a significant background at low momentum transfers (e.g. <20 nm$^{-1}$, and especially <5 nm$^{-1}$), while more generally any sample environment near to the sample (e.g. diamonds and gaskets and pressure medium for a diamond anvil cells, capillaries or windows for liquids, glue or varnish for solids) can contribute both inelastic (phonons) and elastic scattering (Bragg or SAXS) which can lead to serious problems with background. This must be carefully considered on a case-by-case basis. Often, sample cells are made with single crystal windows (diamond, sapphire, silicon) as they reduce the SAXS and produce only a well-defined phonon background. The high sound-speeds and generally high phonon frequencies of the diamond and sapphire can be especially advantageous as the overlap with interesting frequencies in other materials can be small.

## 6. Phonon Calculations

Calculations are required in order to know where to measure, and to understand the results. At the very simplest levels, phonons can be calculated using a Born-von-Karman (BvK) force constant model, where the atoms are considered as masses with the force-constants (springs) between pairs of atoms being the free parameters. At a higher level of sophistication one may choose one of several different models with more physically direct parameters (see, e.g. [12,13] and references therein) of which the most well known is probably a shell model. In other cases, one can do some sort of ab-initio modeling, based on density functional methods, or molecular dynamics. Any of these models is immediately useful to gain a first look at the symmetry properties of the modes in a material which provide, of course, many constraints on the motion of atoms. In full generality, these can be investigated using group theory (see, for example, the web-based applications at the bilbao crystallographic server [39]), but, to some extent, any model will begin to reflect these and allow first approximation to understanding the modes of the sample. Such models can often be built beginning with results from similar materials in the literature. Available shell model codes include OpenPhonon [40], Unisoft [41], and GULP [42], of which the first two have been shown to give mostly consistent results. Practically, in many cases, these models can be reduced to a BvK set of spring-constants (the real-space force constant matrix), but the different models have the advantage of being in terms of more physically transparent quantities.

Density functional (DFT) methods offer several approaches to the phonons, and in many cases, can do a fantastic job of predicting phonon behavior. The phonon calculations, however, consume significant CPU time. Linear response theory (which assumes a pseudo-harmonic model) is perhaps the least computationally intensive method, though super-cell methods also offer some advantages. Generally available codes include PWscf [43] and ABINIT[44]. Others are available based on collaborative or financial agreements. Finally, one notes that molecular dynamics codes can also be used to calculate phonons, and may have some advantages for determination of phonon lifetimes, but, to our experience, remain the least used of the available methods for samples measured using IXS.

A useful (portable) form for the result of phonon calculations is the real-space force constant matrix, generated by interpolating results of calculations at a series of points in reciprocal space. This is essentially a list of 3x3 matrices corresponding to the Born-von-Karman force constants between pairs of atoms. This may be extended over many pairs of atoms (say, out to 5 or 10 nearest neighbors) to improve accuracy. The real-space matrix can be used to calculate the phonon energy and polarization at any momentum transfer. Phonon line-widths are more complex. Linear response theory can be used to estimate the lowest order line-width from electron-phonon coupling, while phonon-phonon scattering, anharmonicity, often requires a special effort for specific momentum transfers. In some cases, it is possible to imbed line-width information into a matrix form analogous to the real-space force constant matrix [45].

## 7. Examples

Superconductors are probably the single largest class of crystalline materials investigated using IXS. This is because, in the conventional model of superconductivity, the phonons promote the formation of cooper pairs. The subject also has technological importance, both for creation of new devices, and for issues such as distribution of electrical power. Finally, the continued mysteries surrounding the mechanism of high-$T_c$ superconductivity in cuprates, and, most recently the iron pnictide materials, is a source of challenging scientific questions. In this context, IXS offers access to small samples, and the possibility to investigate high-energy modes with good resolution in both energy and momentum space. Superconductors investigated at BL35XU include $MgB_2$ [26], $HgBa_2CuO_{4+\delta}$ (Hg1201) [46], $La_{2-x}Sr_xCuO_4$ (LSCO) [47], Boron-doped diamond [48], CaAlSi [34], $YBa_2Cu_3O_{7-\delta}$ (YBCO)[38], $Bi_2Sr_{1.6}La_{0.4}Cu_2O_{6+\delta}$ (Bi2201) [49] and FeAs materials [33], to name a few. Here we choose 3 samples that show different features: CaAlSi, as an example of soft-mode driven superconductivity, provides a nice contrast to the case of $MgB_2$ that has been extensively discussed previously [26,23,37]. Precision investigations of a phonon anomaly in the high-$T_c$s gives an example of some of the advantages of IXS over INS, and finally we use to the case of FeAs to show that density of states can generate useful immediate information.

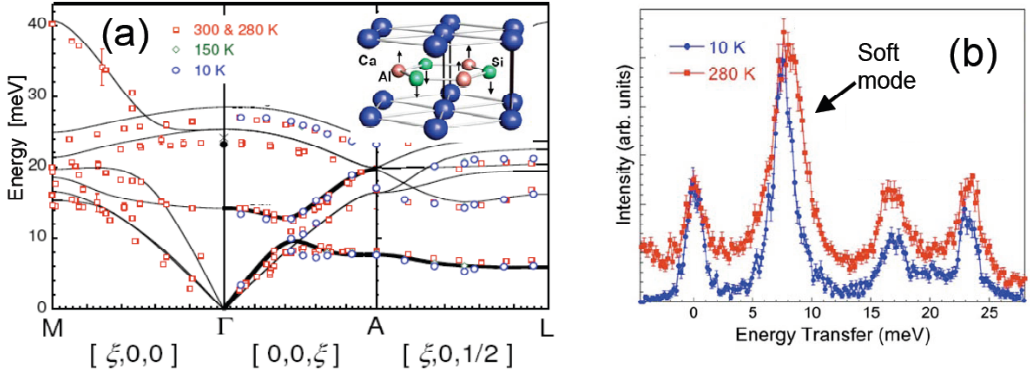

**Fig. 2.** Phonons in CaAlSi (after Ref. [34]). (a) shows the dispersion compared against ab-inito calculation in a local density approximation (LDA) while (b) shows measured spectra at the A point. The inset in (a) shows the motions of the atoms in the soft mode. Please see the text for details.

CaAlSi is a superconductor with a structure similar to that of $MgB_2$ – it has hexagonal planes of Al and Si separated by Ca (see figure 2). The driving mechanism of superconductivity in this case, however, is a low-energy mode resulting from a structural instability related to the buckling of the Al-Si planes. Thus for example, there are several different structures depending on the ordering and buckling of the AlSi planes [50]. Unsurprisingly, then, the relevant phonon mode, the buckling mode of the AlSi planes, is unusually soft. This is an interesting contrast to the case of $MgB_2$, where it is the electron-phonon coupling that drives the phonon softening. The figure shows the measured dispersion compared against ab-initio calculations, and spectra at the "A" point with the soft mode clearly visible. The large line-width suggests significant anharmonicity at room temperature, and the increased background is probably due to multi-phonon contributions. At low temperatures the line-width of the soft mode agrees well with expectations from ab-inito calculations for electron-phonon coupling [34].

The copper oxides remain one of the outstanding interesting problems in correlated materials. Phonons were initially considered as insufficient to drive such high-temperature superconductivity, but, now there is, perhaps, a greater appreciation for a phonon role in the properties of these materials, and in their superconductivity. From the point of view of the phonon spectra, there are several anomalies, of which the largest and, possibly, the most well known is the "bond-stretching anomaly" associated with the softening and, more-so, the large line-width of the copper-oxygen bond-stretching mode at about half way to the zone boundary – a momentum transfer, that, for example, might correlate with dynamic stripe order. Figure 3 shows this anomaly in $La_{1.48}Nd_{0.4}Sr_{0.12}CuO_4$:

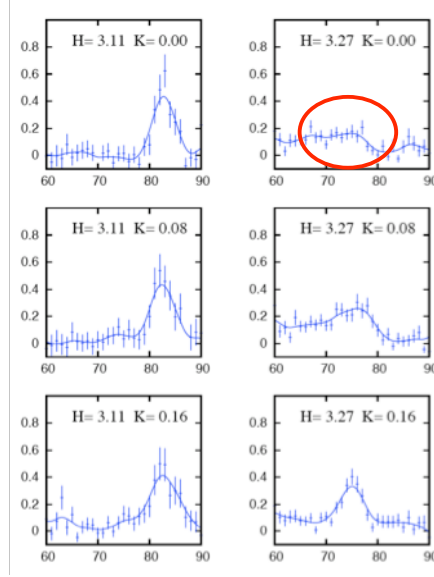

**Fig. 3.** IXS spectra of $La_{1.48}Nd_{0.4}Sr_{0.12}CuO_4$ (after ref. [51]). The circle indicates the position of the bond-stretching anomaly. Please see text for discussion. The horizontal axis is energy transfer in meV. Lines are to guide the eye. The horizontal scale is energy transfer in meV and the vertical scale is intensity. Note L=0 for all spectra.

the phonon mode becomes extremely broad and nearly invisible in one very specific region of momentum space (red circle) [51]. The interesting point about the present work is that it shows a rather sharp momentum dependence of the anomaly, with it being well-localized in both directions in momentum space. This requires a re-interpretation of some previous neutron results [52]. Another notable point is that the spectra in the figure required 2.5 days of data collection into each spectrum to accumulate acceptable data. This highlights the advantages of a 2 dimensional analyzer array (which allowed all the spectra in figure 3 to be collected at the same time), and also the need for increased intensity. Finally, one should note that the energy resolution, ~1.8 meV at >80 meV energy transfer would be difficult to achieve using neutron scattering.

Finally, we show work from a now-ongoing project investigating the new iron pnictide superconductors [33]. These systems have shown superconductivity at temperatures as high as 55K, and like the cuprates, the superconductors are doped systems, with the parents showing magnetic order. Initial calculations suggested these materials were not phonon mediated, however, measurements of the phonon density of states quickly showed that the calculations failed to reproduce the frequencies of the iron modes. This was done using IXS, both by looking at the density of states in powders (fig 4) and also the dispersion of a single crystal (fig 5) [33]. The important points for IXS as a technique here are that (1) it is possible to get information from powder samples, on the one hand, and (2) that single crystal measurements are possible with small samples – in this case about 0.15x0.2x0.02 mm$^3$.

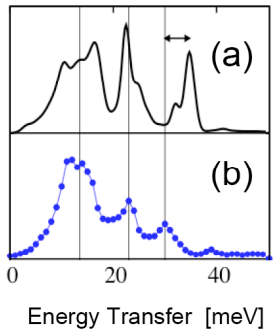

**Fig. 4**. Calculated (a) and measured (b) Pseudo-DOS from LaFeAsO (after [33]).

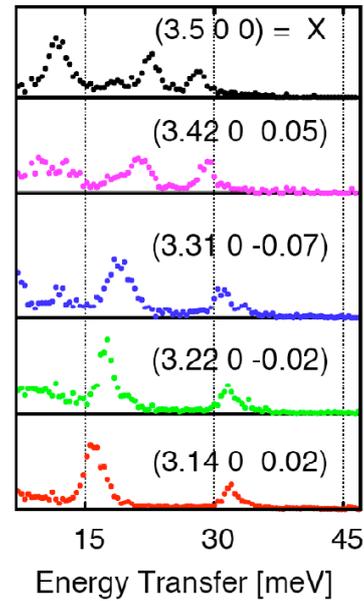

**Fig. 5.** IXS spectra from a single crystal PrFeAsO$_{1-\delta}$ (after [33])

Meanwhile the results suggests that even in the nominally non-magnetic superconductors, the phonons show evidence of magnetism, agreeing better with spin-polarized calculations than with the non-magnetic calculations. One also notes recent evidence of a strong isotope effect on both the magnetic and superconducting transition temperatures [53]. This all seems to be part of an unfolding storey of strong magneto-elastic coupling in these materials.

**7.1 Other Materials**
Here we briefly mention a few of the other types of experiments that have been done with IXS. Ferroelectrics, with the coupling of lattice and electronic polarization, continue to be an interesting topic: there is work on relaxor materials [54] and multi-ferroics[55]. Likewise, charge-density-wave (CDW) materials are of interest, and there are a couple unpublished data sets [56][57]. Meanwhile, localized modes in some materials, including skutterudites [58], are under investigation due to their possible relevance to thermoelectricity and to f-electron physics. Meanwhile several studies are also under way, and have been carried out [59], to investigate sound velocities under high-pressure conditions by measuring the dispersion of acoustic modes using with powders or single crystals in diamond anvil cells. In principle, this technique offers rather precise results for sound velocities [60] that are difficult to measure by other methods. There is also effort to extend studies into a surface sensitive geometry, both in Japan and at other facilities [61].

## 8. Concluding Comments
The field of IXS has grown very quickly over the past decade as new instruments have come on line. With the easy access to small samples and a relatively simple setup, it offers users a convenient, even easy, probe of atomic dynamics in materials. Additionally, there is interest stemming from the increased focus of materials science on complex and correlated materials for technological applications. Here the phonons are an important part of an interconnected set of properties that make materials technologically useful. Thus there is increasing drive to measure, and increased interest in, atomic dynamics. Furthermore, the ability to measure tiny, ~microgram, crystals with relative ease, making it possible to measure a new material very quickly after it is first synthesized, provides added impetus to investigate phonons.  When backed up by increased availability of modeling codes, and increase computational speed, measurements of phonon dispersion are poised to move from an interesting, if slightly esoteric field, to a more mainstream tool of solid-state-physicists and materials scientists.  The largest limitations for IXS remain available beamtime and flux on the sample.  However, all instruments are continually upgrading and becoming better.  Also, there are two new instruments that are now funded for high resolution IXS, including a new facility in Japan, BL43LXU.  In particular, among other improvements, BL43LXU is expected to provide about one order of magnitude more flux on the sample than is now available, so that a new generation of experiments will become possible.

## Acknowledgements
We are indebted to Drs. H. Uchiyama, Y. Kohmura, K. Sawada, and D. Ishikawa for kind and useful discussion.  We thank Ms. N. Kosami of the RIKEN SPring-8 Center for patient work on the Japanese manuscript.